\documentclass[prb,aps,12pt,nofootinbib,tightenlines]{revtex4}
\usepackage{epsf}

\def\half{{1\over 2}}
\def\be{\begin{equation}}
\def\ee{\end{equation}}
\def\ba{\begin{eqnarray*}}
\def\ea{\end{eqnarray*}}
\def\dt{\Delta \tau}
\def\tY{\tilde{Y}}
\def\btheta{\bar{\theta}}
\begin{document}
\title{Quantum phase slips in a confined geometry}
\author{S. Khlebnikov}
\affiliation{Department of Physics, Purdue University, West Lafayette,
Indiana 47907, USA}

\begin{abstract}
We consider tunneling of vortices across a superconducting film that is
both narrow and short (and connected to bulk superconducting 
leads at the ends). We find that in the superconducting state 
the resistance, at low values of 
the temperature ($T$) and current, does not follow the power-law dependence 
on $T$ characteristic of longer samples but is exponential in $1/T$. 
The coefficient of $1/T$ in 
the exponent depends on the length or, equivalently, the total normal-state 
resistance of the sample. These conclusions persist in the one-dimensional 
limit, which is similar to the problem of quantum phase slips in an
ultra-narrow short wire. 

\end{abstract}

\maketitle

\section{Introduction}
Vortex tunneling across narrow thin films is a quantum effect that limits 
superconductivity in these systems. It does not rely on any vortices preexisting 
in the sample (due, e.g., to temperature or a magnetic field) but occurs as 
a result of quantum 
fluctuations---either a vortex entering the sample from the outside or 
a virtual vortex-antivortex pair created in the sample.
A supercurrent (if present) exerts a force on the vortex and,
for certain types of vortex motion,
the total work done by this force is nonzero. That means that an amount of energy 
is taken away from the supercurrent, i.e., there is dissipation---an electrical 
resistance.

Frequently, one considers vortex tunneling
at a strong enough current, so that the vortex can nucleate close to the boundary:
the closer it nucleates, the shorter is the distance it has to tunnel, and the
larger is the tunneling rate.\cite{Glazman&Fogel,Tafuri&al} 
However, for very narrow samples (starting perhaps with a few tens of nm),
the rate may remain substantial even 
at weak currents, when the vortex tunnels the entire width. 
This is the case we consider here, motivated in part by its similarity
to quantum phase slips in a genuinely one-dimensional (1d) wire, a topic of much 
current experimental 
research.\cite{Giordano,Giordano&Schuler,Giordano2,Bezryadin&al,Lau&al,Tian&al,
Zgirski&al,Rogachev&al,Altomare&al,Bollinger&al,Bollinger&al2007}

While the core of a vortex can for our purposes be considered point-like, 
it is important to take into account the long-range disturbance that vortex
motion produces. Low-dimensional superconductors support a gapless
plasmon mode,\cite{Kulik,Mooij&Schon} which in the presence of a nearby 
ground plate (the case considered here) has an acoustic dispersion law, 
with a speed $c_0$. This leads to a new length
scale, $l_p = \hbar c_0 /T$ ($T$ is the temperature), 
to which other length scales can be compared. 
In particular, one can distinguish between long wires, those of length 
$L \gg l_p$, for which the final state of plasmons does not depend on 
the boundary conditions at the ends, and short ones, $L \ll l_p$, for which 
it may.

For long 1d wires, plasmon production has been shown theoretically
to significantly affect the tunneling 
rate.\cite{Zaikin&al,Khlebnikov,diso} In the presence of disorder, it leads
to a power-law current-voltage dependence in the superconducting state
at low temperatures,\cite{diso}
$V(I) \propto I^\alpha$ (a state is superconducting if $\alpha > 1$).
Here, we consider the opposite, short-wire,
limit.\footnote{We will often use the system of units with $\hbar=c_0=1$, 
in which the short-wire condition is simply $LT \ll 1$.}
This choice is motivated by a puzzle in the existing experimental results:
while in a long wire the observed nonlinear $V(I)$ curve is indeed well 
described by a power law,\cite{Altomare&al} in short wires, 
no power law in current or in temperature has
been detected.\cite{Rogachev&al} Our results provide an explanation for that.

A general theory that allows one to compute the effect of plasmons on vortex
tunneling for samples of various sizes can be constructed
along the following lines. As well known (and reviewed, for example, in Ref. 
\onlinecite{Wen&Zee}), two-dimensional superfluids have a dual description, 
in which vortices are viewed as charges and plasmons as ``photons'', so that
the theory maps onto planar quantum electrodynamics (QED).
We present a derivation of this, based on a path-integral identity, in Sect. 
\ref{sect:dual}.

The version of QED that we use in this paper is ``quenched'', 
in the sense that it considers only a single vortex and 
neglects interaction with additional vortices that may 
tunnel nearby. We expect this to be
a good approximation as long as one stays away from a superconductor-insulator
transition. (For 1d wires, a mean-field-type theory that takes into account 
interactions between quantum phase slips has been recently proposed in 
Ref. \onlinecite{Meidan&al}.) In the quenched limit, 
the computation of the plasmon action amounts essentially to a Euclidean
(imaginary time) version of the classical radiation theory. (The imaginary time 
appears since we are considering a tunneling process.) An important aspect of the
theory is formulation of the boundary conditions at the ends of the sample.
This is described in Sect. \ref{sect:rad}, and the solution for the plasmon
field is given in Sect. \ref{sect:sol}.

Our final result is that, in the superconducting state of
narrow short wires connected to bulk superconducting 
leads, the resistance due to vortex tunneling, at small temperatures and currents,
is no longer a power law but an exponential in $1/T$. We trace this stronger
suppression to
a large gradient energy that the system must have already when it enters 
the classically forbidden region. 
The coefficient of $1/T$ in the exponent depends inversely on the length $L$.

We wish to reiterate that this result applies only in the superconducting 
state, where the tunneling events (instantons) are rare, and does not preclude 
the possibility of a transition to an insulating state at larger tunneling
rates.

The tunneling process we consider here is in addition to and competes with
the classical, over-barrier process. The latter is an analog of a thermally
activated phase slip in the 1d case.\cite{Little,LA,MH} As we will see, despite
the above-mentioned suppression, the rate of vortex tunneling is exponentially 
larger than the rate of thermal activation over a broad range of parameters.
We hope that
the difference in both the magnitudes of the resistance and its dependence on 
the length will allow one 
to distinguish between the two effects experimentally.

We should also note the difference in the starting points 
for the theories of these two effects. 
The energy of the LAMH saddle point\cite{LA,MH} is due to the phase
slip core (and consequently depends strongly on the Ginzburg-Landau coherence
length $\xi$). In contrast, in our case, the activation energy is that of the
initial tunneling state, which lies far from the top of the potential barrier.
Depletion of the order parameter in this state is still small, and we 
can use the phase-only theory, in which vortex cores are essentially point-like.

\section{Duality map} \label{sect:dual}
The Lagrangian density of the phase-only theory that describes a superconducting 
film in the presence of a nearby ground plate is
\be
{\cal L} = \frac{1}{2g} (\partial_t \theta)^2 - \half K_s (\nabla \theta)^2 
=  \half K_s \partial^\mu \theta \partial_\mu \theta
\; .
\label{L}
\ee
In the second equality here we have switched to the notation of special relativity,
by using instead of time $t$ the coordinate $x^0 = c_0 t$, where
$c_0 = \sqrt{g K_s}$ is the plasmon speed. (In addition, $x^1 \equiv x$ and 
$x^2 \equiv y$.) In what follows, we choose units of
length and time so that $c_0 = 1$. Greek indices take values $0,1,2$, and summation
over a repeated index is implied.

The field $\theta$ is the phase of the order parameter, but no assumption is 
made about the existence of long-range order, i.e., we do no require the
expectation value of $\exp(i\theta)$ to be nonzero. All that is required for 
superfluidity is that the stiffness $K_s$ renormalizes to a nonzero
value in the infrared.

If $\theta$ were a single-valued smooth function of $(x,y)$, plasmons would be 
the only excitations in the system, and the theory would be completely Gaussian. 
To describe vortices, we allow $\theta$ to be multivalued. 
Alternatively, we could make it discontinuous by drawing explicit branch cuts 
at vortex positions, but we will be using the first approach. Then, 
$\nabla \theta$ is smooth outside vortex cores.

The phase-only description (\ref{L}) does not resolve vortex cores, 
so a short-scale
cutoff of order of the Ginzburg-Landau coherence length $\xi$ is implied.

Let us remark on the issue of gauge
invariance and the apparent absence of electromagnetic fields from (\ref{L}). 
In thin films, magnetic field of a vortex extends over a large area,
determined by the transverse screening length $\lambda_\perp$.\cite{Pearl}
If $\lambda_\perp$ (which in thin films can 
exceed 100 $\mu$m) is much larger than the 
smallest dimension of the film, the magnetic field can be neglected,
and it is possible to choose a gauge such that the vector potential
${\bf A}$ is close to zero. The remaining gauge freedom can be used to make 
the scalar
potential $A_0$ go to zero away from the film. This fixes the gauge completely, and
if $\theta$ denotes the phase in this gauge it is in effect gauge-invariant. 
Integrating out $A_0$ produces a capacitive term, 
which eventually becomes the first term in (\ref{L}).

We have not included in (\ref{L})
a ``topological'' term, proportional to $\partial_t \theta$, that gives rise to 
the Magnus force on a vortex. 
We consider films that have significant amounts of disorder, and in disordered 
superconductors the Magnus force is small.\cite{magnus}

Eq. (\ref{L}) allows one to describe dissipation of energy of vortex motion into
plasmons, but does not include any dissipative mechanisms related to normal
electrons at the vortex cores. This is justified in the limit of strong disorder, 
since transfer of
energy to the normal component in this case is inhibited by the short electron
mean-free path.

Because in the presence of vortices $\theta$ is multivalued, the expression
\be
J^\mu = \frac{1}{2\pi}
\epsilon^{\mu\nu\lambda} \partial_\nu \partial_\lambda \theta \; ,
\label{J}
\ee
where $\epsilon^{\mu\nu\lambda}$ is the unit antisymmetric tensor, is nonzero and
indeed is the vortex current ($J^0$ is the vortex density).
Setting
\be
q_\lambda \equiv \partial_\lambda \theta \; ,
\label{q}
\ee
we can write the current (\ref{J}) as
\be
J^\mu = \frac{1}{2\pi}
\epsilon^{\mu\nu\lambda} \partial_\nu q_\lambda \; .
\label{J2}
\ee
The equation of motion following from (\ref{L}) is 
\be
\partial^\lambda \partial_\lambda \theta = \partial^\lambda q_\lambda = 0 \; .
\label{eqm}
\ee
[In the presence of a topological term, the temporal component of (\ref{q}) is
replaced by $q_0 = \partial_0 \theta + {\rm const.}$, and eq. (\ref{sp2}) below
is modified accordingly. The expression (\ref{J2})
for the current and the equation of motion (\ref{eqm}) are both unaffected.]

In what follows we restrict our attention to configurations satisfying the equation
of motion (\ref{eqm}). In the real-time version of the theory, they
describe motion of an arbitrary number of real vortices in the presence of supercurrents 
and plasmon waves. In the Euclidean (imaginary time) version, to which we turn shortly,
solutions to the equations of motion will determine the most probable tunneling paths
(instantons) responsible for the quantum decay of supercurrents.

The real-time action corresponding to the Lagrangian density (\ref{L}) is
\be
S = \half K_s \int \partial^\mu \theta \partial_\mu \theta d^3 x =
\half K_s \int q^\mu q_\mu d^3 x \; ,
\label{S}
\ee
where $d^3 x = dx^0 dx^1 dx^2$. If $q_\mu$ is a solution of the equation of motion 
(\ref{eqm}), there is a path-integral identity---the duality map:
\be
e^{iS}
= \int {\cal D} f_{\mu\nu} {\cal D} \lambda 
\exp  \int i K_s \left\{ -\frac{1}{4} f^{\mu\nu} f_{\mu\nu}
- \half \epsilon^{\mu\nu\rho} f_{\mu\nu} q_\rho 
+ \half \lambda \epsilon^{\mu\nu\rho} \partial_\rho f_{\mu\nu} \right\} d^3 x \; ,
\label{map}
\ee
where $f_{\mu\nu}$ is antisymmetric in $\mu$, $\nu$ and is subject to the boundary
condition
\be
\left. \epsilon^{\mu\nu\rho} f_{\mu\nu} n_\rho \right|_b = - 2 n^\rho q_\rho \; ;
\label{bc}
\ee
$n_\rho$ is the normal to the boundary ($b$) of the spacetime volume.

Because the path integral in (\ref{map}) is Gaussian, the map can be verified 
directly. First, note that the path integral over $\lambda$
enforces the Bianchi identity
\be
\epsilon^{\mu\nu\rho} \partial_\rho f_{\mu\nu} = 0 \; .
\label{bianchi}
\ee
Since in (\ref{map}) $f_{\mu\nu}$ is an independent variable and not (yet) a curl
of some gauge field, (\ref{bianchi}) is not really an identity but an 
independent equation of motion; however, we keep the familiar term.

Next, integrating over  $f_{\mu\nu}$ amounts to solving the saddle-point equation
\be
f^{\mu\nu} + \epsilon^{\mu\nu\rho} (q_\rho + \partial_\rho \lambda) = 0 \; .
\label{sp}
\ee
Taking curl of this and using eqs. (\ref{bianchi}) and (\ref{eqm}), we obtain 
$\partial^\rho \partial_\rho \lambda = 0$. On the other hand, applying the boundary
condition (\ref{bc}) in eq.~(\ref{sp}) tells
us that the normal derivative of $\lambda$ at the boundary is zero. For a spacetime
volume of a simple shape, these conditions are sufficient to reduce 
$\lambda$ to a constant, which then drops out of (\ref{sp}). Eq. (\ref{sp}) becomes
\be
f^{\mu\nu} = - \epsilon^{\mu\nu\rho} q_\rho =  - \epsilon^{\mu\nu\rho} \partial_\rho
\theta \; .
\label{sp2}
\ee
Substituting this back in (\ref{map}) we confirm that the original action (\ref{S}) 
is recovered. In terms of the saddle-point value (\ref{sp2}), this action can be
written as
\be
S = \frac{1}{4} K_s \int f^{\mu\nu} f_{\mu\nu} d^3 x \; .
\label{sp_action}
\ee

Now, let us take an alternative (``dual'') view and solve the identity (\ref{bianchi})
explicitly by introducing a new gauge field $a_\mu$ (quite 
distinct from the electromagnetic potential $A_{\mu}$):
\be
f_{\mu\nu} = \partial_\mu a_\nu - \partial_\nu a_\mu \; .
\label{f_munu}
\ee
Differentiating (\ref{sp2}), we obtain Maxwell equations for this field:
\be
\partial_\nu f^{\mu\nu} = - 2\pi J^\mu \; ,
\label{maxwell}
\ee
where $J^\mu$ is the vortex current (\ref{J2}). The usefulness of this dual view is 
that it allows one to determine $f_{\mu\nu}$, and the corresponding action, for any
prescribed motion of vortices. Using (\ref{sp2}), one can then find
the derivatives of the phase $\theta$.

The dual theory also allows one to take into account the backreaction
of produced ``photons'' on the vortex motion, by extremizing the total action 
with respect to that motion itself. In the problem to which we apply this theory
here, backreaction will determine the region where the vortex prefers to 
tunnel. (This region turns out to be the middle of the wire---in
contrast to
the resistively shunted 1d case, where quantum phase slips occur
preferentially near the ends.\cite{Buchler&al})

To any solution of eq. (\ref{maxwell}) we can add a homogeneous solution---a
static uniform ``electric'' field $f_{0i} = {\rm const}$. According
to (\ref{sp2}), such a field corresponds to a static uniform supercurrent. We adopt 
the convention in which the supercurrent density is measured in units of $-2|e|$
($e$ is the electron charge), 
i.e., is given by $K_s \nabla \theta$. Then, for example, a current in the positive 
$x$ direction
corresponds to an ``electric'' field in the negative $y$ direction, and if a vortex
moves that way, far from any boundaries, the work done on it by the current 
will be positive. In the presence of boundaries, a nontrivial change of $\theta$
at a boundary may give an additional contribution to the total work.

\section{Radiation theory} \label{sect:rad}
To describe tunneling, we switch to the Euclidean time, via
\ba
\tau & = & i t \; , \\
a_4 & = & - i a_0 \; , \\
J^4 & = & i J^0 \; .
\ea
The relations (\ref{sp2}) take the form
\begin{eqnarray}
-i \partial_4 \theta & = &  f_{xy}  \equiv B \; ,
\label{B} \\
i \partial_y \theta & = & f_{x 4}  \equiv  E 
\; , \label{E} \\
- i \partial_x  \theta  & = &   f_{y 4} \equiv F 
\; , \label{F}
\end{eqnarray}
where we have introduced shorthands $B$, $E$, and $F$, which will be much used in 
what follows.

The Euclidean counterpart of the action (\ref{sp_action}) is
\be
S_E = - \half K_s \int dx d y d\tau \left(
B^2 +  E^2 +  F^2 \right) \; .
\label{SE}
\ee
Note that, as a result of the transition to the Euclidean time, the relations 
of all three components of $f_{\mu\nu}$ to the derivatives of $\theta$ have acquired 
factors of $i$. As a consequence, on the instanton solution, $B$, $E$, and $F$ will 
all be purely imaginary, and the action (\ref{SE}) will be positive. 
This is in agreement 
with the a priori expectation that coupling to plasmons should suppress vortex 
tunneling (the suppression factor is $e^{-S_E}$) since, as it tunnels, the vortex 
has to drag the plasmon subsystem with it. 

Maxwell equations (\ref{maxwell}) in these notations have the form
\begin{eqnarray}
-\partial_4 E -  \partial_y B & = & 2\pi J^x \; , \label{eq_noF} \\
- \partial_4 F + \partial_x B & = & 2\pi J^y \; , \label{eqF1} \\
\partial_x E +  \partial_y F & = & 2\pi J^4 \; , \label{eqF2}
\end{eqnarray}
while the Bianchi identity reads
\[
\partial_x F = - \partial_4 B +  \partial_y E \; .
\]
Using these together, one can obtain independent wave equations for $B$, $E$, and 
$F$.
All that is left to choose, then, is a suitable form of the vortex
current and the boundary conditions for the fields.

We consider the theory on a rectangular strip of length $L$ ($0 \leq x \leq L$) 
and width $w$ ($0\leq y \leq w$) and assume that the vortex motion is purely
transverse: $J^x = 0$. Then, the wave equations are
\begin{eqnarray}
\partial_4^2 B + \nabla^2 B & = &  2 \pi \partial_x J^y \; , \label{eqB} \\
\partial_4^2 E + \nabla^2 E & = &  2 \pi \partial_x J^4 \; . \label{eqE}
\end{eqnarray}
Once solutions to these are obtained, the solution for $F$ can be found from
eq. (\ref{eqF1}) or (\ref{eqF2}), except for the static uniform component. 
The latter is the static uniform ``electric'' field mentioned at the end of
Sect. \ref{sect:dual}. It corresponds to a steady supercurrent
in the $x$ direction, and that supercurrent can in principle have any value. 
Eventually, this component of $F$ will be determined by the properties
of the metastable state from which the system tunnels.

The remaining (nonzero) components of the vortex current are
\begin{eqnarray}
J^y & = & i \partial_\tau  Y \delta(x -X) \delta(y - Y(\tau))  \label{Jy} \; , \\
J^4 & = & i  \delta(x -X) \delta(y - Y(\tau)) \label{J4} \; ,
\end{eqnarray}
where $Y(\tau)$ is the transverse position of the vortex. At a finite temperature $T$, 
$Y(\tau)$ must be periodic in $\tau$ with period $\beta = 1/T$.

\begin{figure}
\leavevmode\epsfysize=2in \epsfbox{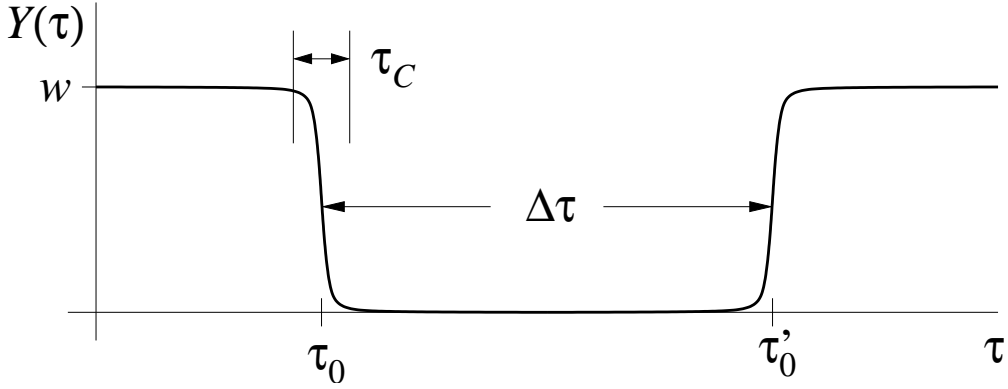}
\vspace*{0.2in}
\caption{The vortex's transverse  position as a function of the Euclidean time.
$\Delta \tau$ denotes the instanton-antiinstanton separation, and $\tau_C$ the
duration of an instanton. In the text, we also use the rescaled variable 
$\tY = Y /w$, where $w$ is the wire's width.}
\label{fig:Y}
\end{figure}

If the vortex could nucleate inside the strip, $Y(\tau)$ would start at the upper 
edge, $Y = w$, move down to
$Y = Y_{\rm nucl}$, the nucleation point, and then back to $Y = w$. 
This would form a ``bounce''.\cite{Coleman}
As we already
noted, though, here we consider only supercurrents that are small enough for
the vortex to have to tunnel the entire width $w$. Then, the relevant 
configuration is an instanton-antiinstanton (IA) pair: a vortex tunneling across the
strip around time $\tau= \tau_0$, plus an antivortex 
(or a vortex moving in the opposite direction) 
tunneling around $\tau = \tau_0'$. A representative history of $Y(\tau)$ is 
shown in fig. \ref{fig:Y}. 

Turning to the boundary conditions (b.c.), we note that,
since there is no current through
the edges of the strip, $B$ and $E$ satisfy, respectively, the Neumann and 
Dirichlet boundary conditions at $y=0$ and $w$. 
We can then define Fourier transforms with respect to $y$ and $\tau$:
\ba
B(x,y; \tau) & = & T \sum_{ln} e^{-i \Omega_n \tau} \psi_l(y) B_{ln}(x) 
\; , \\
E(x,y; \tau) & = & T \sum_{ln} e^{-i \Omega_n \tau} \chi_l(y) E_{ln}(x) 
\; ,
\ea
where $\psi_l(y)=\cos(\pi l y /w)$ and
$\chi_l(y)=\sin (\pi l y /w)$, and the sum over $l$ starts from $l=0$
in the first case and from $l=1$ in the second; $\Omega_n = 2\pi n T$ are
the Matsubara frequencies. In either case, the wave operator 
takes the form 
\[
\partial_4^2 + \nabla^2 \to \partial_x^2 - k_{ln}^2 \; ,
\]
with
\be
k_{ln}^2 = \frac{\pi^2 l^2}{w^2} + \Omega_n^2 \; .
\label{kln}
\ee
The action (\ref{SE}) becomes
\be
S_E = - \half K_s w T \sum_{ln} C_l \int_{0}^{L} dx   \left(
B_{ln} B_{l,-n} +  E_{ln} E_{l,-n} +  F_{ln} F_{l,-n}  \right) \; ,
\label{SEF}
\ee
where $C_l=1$ for $l=0$ and $C_l = \half$ otherwise.

To obtain the b.c. at $x = 0,L$, we need to specify how the sample connects to the
outside world. Here, we consider the case when the leads are bulk superconductors.
As a model of those, we use strips of the same width $w$ and with same
parameter $g$ as the wire but of much
larger stiffness, $K_s' \gg K_s$, and length $L' \gg L$.
$L'$ will eventually be taken to infinity.

Consider the interface at $x=L$. Denote the Fourier components of $\theta$ there as
\[
\theta_{ln}(L) \equiv \btheta_{ln} \; .
\]
Then, for any $l$ and $n$ that are not both zero, throughout the lead ($x > L$)
\be
\theta_{ln}(x) = \btheta_{ln} e^{-k'_{ln} (x - L)} \; ,
\label{lead}
\ee
where $k'_{ln} > 0$ and is given by (\ref{kln}) with $\Omega_n^2$ replaced by
$\Omega_n^2 / g K_s'$.
Substituting (\ref{lead}) into the action of the lead, we obtain a contribution to
the effective action of $\btheta_{ln}$:
\[
S'_E = 
\half K_s' w T \sum_{k'_{ln} > 0} C_l k_{ln}' \btheta_{ln} \btheta_{l,-n} \; .
\]
When we extremize the total action with respect to $\btheta_{ln}$, 
this term gives $\btheta_{l,-n}$ with a coefficient that
grows at least as $\sqrt{K_s'}$ at large $K_s'$. As a result, at large $K_s'$,
$\btheta_{ln}$ are close to zero. 
The same applies at the other interface, at $x =0$.

We conclude that, in the case of bulk superconducting leads, both $B$ and
$E$ satisfy the Dirichlet b.c. at either end:
\begin{eqnarray}
B_{ln}(0) = B_{ln}(L) & = & 0 \; , \label{bcB} \\
E_{ln}(0) = E_{ln}(L) & = & 0 \label{bcE}
\end{eqnarray}
($k_{ln} > 0$). The absence of $k_{ln} = 0$ from these conditions is inconsequential,
since neither $B$ nor $E$ has an $l=n=0$ mode (only $F$ does). 

Note that the condition (\ref{bcB}) is not satisfied by the trial instanton
configurations considered (for the 1d case) in Ref. \onlinecite{Golubev&Zaikin}. This
explains the difference in the final results: Ref. \onlinecite{Golubev&Zaikin} finds
that the phase slip rate remains finite at $T\to 0$, while we find that it is
exponential in $1/T$.

We now proceed to solving the wave equations for $B$ and $E$, and determining $F$
and the Euclidean action.

\section{Solution for the plasmon} \label{sect:sol}
Since the current components (\ref{Jy}) and (\ref{J4}) have the same $x$-dependence,
both $B$ and $E$ can be expressed through solutions to the equation
\be
-\partial_x^2 f_{ln} + k_{ln}^2 f_{ln} = \partial_x \delta(x - X) 
\label{f}
\ee
($k_{ln} > 0$) with the Dirichlet b.c. $f_{ln}(0) = f_{ln}(L) = 0$. 
These solutions are readily found:
\be
f_{ln}(x) = \frac{1}{\sinh(k_{ln} L)} \left\{ \begin{tabular}{ll}
$\cosh[k_{ln} (L - X)] \sinh (k_{ln} x) \; ,$ & $~~~~~x < X \; ,$ \\
$- \cosh(k_{ln} X) \sinh[k_{ln}(L - x)] \; ,$ &  $~~~~~x > X \; ,$
\end{tabular} \right.
\label{ff}
\ee
although, as we will see, many useful conclusions can be drawn even without using 
this explicit form.

From now on, we assume that the width $w$ is much smaller than the length $L$ 
(i.e., $w \ll L$) and restrict attention to the $l=0$ ($y$-independent) modes, 
which are the only ones that are potentially infrared sensitive. 
For tunneling paths
such as the one shown in fig. \ref{fig:Y}, this makes our problem similar to
the problem of a quantum phase slip in a genuinely 1d wire.

The field $E$ has no $l=0$ modes, so it drops out of the subsequent
discussion. The other fields are
\begin{eqnarray}
B_{0n}(x) & = & - 2\pi i (\partial_\tau \tY)_n f_{0n}(x) \; , \label{solB} \\
F_{0n}(x) & = & \frac{2\pi}{\Omega_n}   (\partial_\tau \tY)_n 
[\partial_x f_{0n}(x) + \delta(x - X) ] \; , ~~~~~~~~~~~ n \neq 0 \; , \label{solF}
\end{eqnarray}
where $\tY$ is the rescaled transverse coordinate, $\tY = Y / w$, 
and $(\partial_\tau \tY)_n$ is the Fourier transform of $\partial_\tau \tY$:
\be
(\partial_\tau \tY)_n = \int_0^\beta d\tau
\partial_\tau \tY e^{i \Omega_n \tau} = - i \Omega_n \tY_n \; .
\label{Yn}
\ee
The $n=0$ component of $B$ is zero, while that of $F$ is the static uniform 
component that should be determined 
from the properties (the winding number) of the metastable state.

Let us take up the latter task first. In the wire, 
set $F_{00} = - i I / K_s w T$, where $I$ needs to be determined. 
Then, in the leads, $F_{00} = - i I/ K_s' w T$. The total winding number is
\be
N(\tau) = i \int_{-L'}^{L+L'} F(x, \tau) dx = 
\frac{I}{w} \left( \frac{L}{K_s} + \frac{2L'}{K_s'} \right) 
+ 2\pi [\tY(\tau)  - T \tY_0] \; .
\label{tot}
\ee
The last term here is due to the $n\neq 0$ modes (\ref{solF}):
\[
i T \sum_{n\neq 0} e^{-i\Omega_n \tau}
\int_0^L F_{0n} (x) dx = 2\pi T \sum_{n\neq 0} e^{-i\Omega_n \tau} \tY_n 
=  2\pi [\tY(\tau)  - T \tY_0] \; .
\]
Note that the sum evaluates not to
$\tY(\tau)$ but to $\tY(\tau)$
without the zero mode. As a result, $N$ at a given time depends on the entire
history of $\tY(\tau)$, in particular, on the value of the IA separation $\dt$.

The initial and final states of tunneling correspond to points midway between 
the instanton and antiinstanton. These are the points at which the system 
enters and leaves the classically forbidden region. If the instanton and
antiinstanton positions are sharply defined, i.e., $\tau_C \ll \dt$ 
(cf. fig. \ref{fig:Y}), the corresponding times are
\ba
\tau_f & = & \half (\tau_0 + \tau_0') \; , \\
\tau_i & = & \tau_f - 1/2T \; .
\ea
The winding numbers at these times can be computed from (\ref{tot}) and compared
to those of ground states with uniform supercurrents. 

In particular, the winding number at $\tau = \tau_i$ is the same as in 
the uniform ground state with supercurrent $I_{gs}$ given by
\be
\frac{I_{gs}}{w} \left( \frac{L}{K_s} + \frac{2L'}{K_s'} \right) = N(\tau_i) \; .
\label{gs}
\ee
Thus, the initial state belongs to the thermal ensemble built near that ground 
state. A similar relation (with a different ground-state current)
applies in the final state, and we find, as expected,
that the instanton describes tunneling between two thermal ensembles
that differ by a $2\pi$ of the winding number.

The duration of an individual instanton, $\tau_C$ in fig. \ref{fig:Y}, is 
determined by the parameters of the sample. Meanwhile, as we
will see, the IA separation $\dt$ is controlled by the values of the temperature 
and current and becomes large when those are small. So, for 
calculating (\ref{tot}), we can approximate $\partial_\tau \tY$ in (\ref{Yn}) as
\be
\partial_\tau \tY \approx 
- \delta(\tau - \tau_0) + \delta(\tau - \tau'_0) \; .
\label{sharp}
\ee
Then, the last term in (\ref{tot}) becomes
\[
\tY(\tau_i) - T \tY_0 = T \dt \; ,
\]
where $\dt = \tau_0' - \tau_0$. In the limit $L' \to \infty$, eq. (\ref{gs}) 
gives
\[
I_{gs} = I +\frac{\pi K_s' w T \dt}{L'} \; .
\]
This relation between $I$ and $I_{gs}$ can be used to learn 
how much action is contained in the $l=n=0$ mode. The
main contribution comes from the leads, where $F_{00} = - i I/ K_s' w T$ and 
$F_{gs,00} = - i I_{gs}/ K_s' w T$. So, the $l=n=0$ term in the action 
(\ref{SEF}), relative to the corresponding term in the ground state, equals
\be
S_0 = \half K_s' w T \int dx ( |F_{00}|^2 - |F_{gs,00}|^2 ) 
= - 2\pi I \Delta \tau + O(1/L') \; .
\label{S0}
\ee
Note that this is the only term in the action that distinguishes between 
direct and reverse processes: it would change sign if we considered an antivortex 
tunneling at $\tau=\tau_0$ (or a vortex tunneling in the opposite, 
positive $y$, direction).

Turning to the $n\neq 0$ modes, given by (\ref{solB}) and (\ref{solF}), we
find that their action is
\[
S_1 = 2 \pi^2 K_s w T \sum_{n\neq 0} \int_0^L dx
\left\{ \Omega_n^2 f_{0n}^2 + [\partial_x f_{0n} + \delta(x - X) ]^2 \right\} 
|\tY_n|^2
\; .
\]
Integrating by parts in the derivative term and using eq. (\ref{f}), we bring 
this to the form
\[
S_1 = 2 \pi^2 K_s w T \sum_{n\neq 0} \left\{ \partial_x f_{0n}(0) + 
\Omega_n^2 \int_0^X f_{0n}(x) dx \right\} |\tY_n|^2 \; .
\]
Using the explicit form of $f_{0n}$, we find for the term in the braces
\[
\partial_x f_{0n}(0) +  \Omega_n^2 \int_0^X f_{0n}(x) = \frac{k_n}{\sinh (k_n L)}
\cosh[k_n (L - X)] \cosh(k_n X) \; ,
\]
where $k_n \equiv |\Omega_n|$. This has a minimum at $X = L/2$ (albeit 
a shallow one at small $k_n$), meaning that the 
vortex prefers to tunnel in the middle of the wire. Setting 
$X = L/2$, we finally obtain
\be
S_1 = \pi^2 K_s w T \sum_{n\neq 0} k_n \coth( k_n L/2) |\tY_n|^2 \; .
\label{S1}
\ee

The approximation (\ref{sharp}) correctly reproduces Fourier components of 
$\partial_\tau \tY$ with $k_n \ll 1/\tau_C$. For our purposes, it will be
sufficient to consider only these $k_n$, so we can make the replacement
\be
|\tY_n|^2 = \frac{2}{\Omega_n^2} [ 1 - \cos(\Omega_n \dt) ]
\label{cos}
\ee
in (\ref{S1}) and cut off the sum in the ultraviolet at $k_n = 1/\tau_C$.

The resulting expression for the Euclidean action, 
$S_E = S_0 + S_1$, is applicable both to short ($LT \ll 1$) and to long
($LT \gg 1$) wires. In the latter case, it reproduces the result of 
Ref. \onlinecite{diso},
obtained by considering phase slips directly in the one-dimensional theory.
In what follows, we restrict attention to the former case.

We now need to determine $\dt$. We begin by considering the action (\ref{S1}) for
different values of it. For
\be
|\dt| \ll 1/T \; ,
\label{range}
\ee
the sum in (\ref{S1}) can be approximated by an integral:
\be
S_1 \approx 2\pi K_s w \int_0^{1/\tau_C} \frac{dk}{k} [ 1 - \cos(k \dt) ]
\coth( k L/2) \; .
\label{int}
\ee
We see that substantial contributions can only come from $k \agt 1/|\dt|$. It is
therefore convenient to separate the range (\ref{range}) into two. For
$|\dt| \ll L$, the cotangent can be replaced with unity, and
we are back to the expression\cite{diso} for the long-wire case. The underlying physics
is that the typical wavenumber of plasmons in the final state is of order $1/|\dt|$,
and when this is much larger than $1/L$ the effect of the boundaries is insignificant.
In the present case, however, this is possible only for relatively large currents.
Indeed, extremizing the total action $S_0 + S_1$ with respect to $\dt$, we obtain
the saddle-point value $\dt = K_s w / I$. For this to be much smaller that $L$, 
we need $I \gg K_s w / L$. Although 
this condition does not look altogether prohibitive, from now on we concentrate 
on the opposite, small-current, regime:
\[
I \ll \frac{K_s w}{L} \; .
\]
Then, there is no saddle point either for $|\dt| \ll L$ or for $|\dt| \sim L$.

As $|\dt|$ increases past $L$---while still obeying (\ref{range})---the dependence
of (\ref{int}) on $|\dt|$ becomes linear, and one can verify that, for small currents,
this again precludes a saddle point. So, we turn to $\dt \sim 1/T$ and 
the general expression (\ref{S1}) for $S_1$.

Eq. (\ref{S1}) has an extremum at $\dt = 1/2T$
(derivative of each term in the sum vanishes individually). 
This extremum is in fact a
maximum, as it should be: the original integration was over real 
$\Delta t = -i \dt$, and a saddle point that is a minimum in the real 
$\Delta t$ direction is a maximum in real $\dt$. The contribution from $S_0$ 
will displace the
maximum from exactly $1/2T$, but for small currents this displacement is small, 
and for our purposes $\dt = 1/2T$ is a good approximation. Substituting it into
eq. (\ref{cos}), we find that only odd $n$ contribute, and $S_1$ becomes, to
logarithmic accuracy,
\be
S_1 = 8 \pi^2 K_s w T \sum_{n = 1,3,\ldots}^{n_C}
\frac{1}{k_n} \coth( k_n L/2)  
\approx \frac{\pi^2 K_s w}{2 L T} + 2\pi K_s w \ln\frac{L}{\tau_C} \; .
\label{S1f}
\ee
where $n_C \sim 1 / \tau_C T$. In what follows, we retain only the first, leading, 
term on the right-hand side. 

The power dissipated by vortex tunneling is given by the energy $2\pi I$ that
a vortex releases from the supercurrent, times the difference between the rates of
the direct and reverse processes: 
\[
P(I) = 2\pi I [{\cal R}_+(I) - {\cal R}_-(I) ] \; .
\]
To the exponential accuracy, ${\cal R}_+(I) \sim \exp[-S_0(I) - S_1]$, and 
${\cal R}_-(I) = {\cal R}_+(-I)$. Expanding in small $I$, we obtain the resistance:
\be
R(T) \sim e^{-S_1} \sim \exp\left( - \frac{\pi^2 K_s w}{2 L T} \right) \; .
\label{res}
\ee

The activated behavior of the resistance can be interpreted by looking at the
gradient of the phase $\theta$ at the entry point of tunneling, $\tau = \tau_i$.
Using the approximation (\ref{cos}) and setting $\dt = 1/2T$, we obtain
\be
\partial_x \theta(x,\tau_i) = i F(x,\tau_i) 
= \frac{I}{K_s w} + 4 \sum_{n=1,3,\ldots} \frac{1}{n}
\sin(\pi n /2) [\partial_x f_{0n}(x) + \delta(x - X) ] \; .
\label{grad}
\ee
For low-frequency modes, those with $0 < k_n \ll 1/L$, 
\[
\partial_x f_{0n}(x) + \delta(x - X) \approx 1/ L \; ,
\]
so the total contribution of these modes to (\ref{grad}) is 
$(\partial_x \theta)_{\rm low} \approx \pi / L$.
Then, the gradient energy contained in these modes is 
\be
{\cal E} = \half K_s w \int_0^L (\partial_x \theta)^2_{\rm low} dx 
\approx  \frac{\pi^2 K_s w}{2 L} 
\; ,
\label{ene}
\ee
in precise accord with (\ref{res}). This means that the main effect suppressing
resistance at low temperature is the population of the initial tunneling state,
due to the large gradient of $\theta$ already required by that time. There is, 
of course, an additional action associated with the tunneling itself, 
but the precise agreement between (\ref{res}) and (\ref{ene}) implies that 
it is only subleading.

We should note that the entirely classical appearance of the exponent in 
(\ref{res}) (it does not require any powers of $\hbar$) does not contradict
it being a consequence of tunneling, rather than
a classical, over-barrier process. Indeed, restoring $\hbar$ and $c_0$
in the short-wire condition, under which (\ref{res}) applies, 
we obtain $T \ll \hbar c_0 / L$, and this cannot
be realized outside of quantum mechanics. 

For comparison, let us list nucleation energies for two purely classical
processes. The nucleation 
energy of a vortex is of order $\pi K_s$ (times a logarithm), 
and that of the $y$-independent saddle point, analogous to the LAMH saddle point 
in 1d wires,\cite{LA,MH} is of order $K_s w / \xi$, where $\xi$ is 
the Ginzburg-Landau coherence length. As long as $w$ and $\xi$ are both
much smaller than $L$, either of these energies is larger than the energy 
(\ref{ene}). 

Due to a relatively large numerical factor ($\pi^2 /2$) in (\ref{ene}), 
the condition $\xi \ll L$, under which the energy (\ref{ene}) is smaller than
the LAMH activation energy, may in fact mean that $L$ must be several times 
larger than $\xi$. Moreover, even if this condition is satisfied at 
low temperatures, it breaks down in a region close to the critical temperature 
$T_c$. The short-wire condition also breaks down near in a region
near $T_c$, since $K_s$ and hence $c_0$ are small there; in that region,
subleading terms in the instanton action become non-negligible.
Outside of these regions, however, vortex tunneling is the dominant 
resistive process.

For a superconductor in the dirty limit, the exponent in (\ref{res}) can be 
expressed entirely in terms of the superconducting gap $\Delta\equiv \Delta(T)$ 
and the normal-state resistance $R_N = \rho L /  w$, where $\rho$ is
the sheet resistivity. Indeed, in this case 
$K_s = (\pi \Delta / 4 e^2 \rho)\tanh(\Delta / 2T)$,\cite{Abrikosov&al} so 
\be
R(T)  \sim \exp\left( - \frac{\pi^2 R_q \Delta }{4 R_N  T} 
\tanh\frac{\Delta}{2T} \right) \; ,
\label{res2}
\ee
where $R_q = \pi/2e^2 = 6.5$ k$\Omega$. This suggests that the value $R_N = R_q$ 
may have a special significance in short wires. Experimentally, it does: 
this is the value near which one observes a superconducting-insulating 
transition.\cite{Bezryadin&al,Bollinger&al2007}

\section{Conclusion}

We have described a general method for calculating the effect of plasmons on vortex
tunneling in superconducting wires and applied it to the limit of small
temperatures and currents, when the vortex has to tunnel the entire width of the
wire. The method is based on a duality
map, through which vortices become charges and plasmons become ``photons''. 
We have found that, if plasmons
cannot easily leave the tunneling region, as is the case when the wire is short
and the leads are bulk superconductors, the suppression of the resistance in the
superconducting state is exponential at low temperatures and expressed
by eq. (\ref{res}).

As the width of the sample is made smaller and approaches the coherence length 
$\xi$, vortex tunneling crosses over to quantum phase slips in a genuinely 
1d geometry.
The restriction to modes independent of $y$ (the transverse coordinate) that 
we made in sect.
\ref{sect:sol} effectively brings us one dimension down, so we expect the 1d case 
to be similar to ours. 

Despite the exponential suppression, the resistance due to (thermally-assisted)
vortex tunneling is 
larger, over a broad range of parameters, than that
due to classical, over-barrier processes, such as motion of a thermally nucleated
vortex-antivortex pair or a thermally-activated phase slip. In addition, it
has a characteristic dependence on the length $L$ of the
sample or, equivalently, on the total normal-state resistance $R_N$, 
cf. eq. (\ref{res2}).
We hope that these features will allow one to distinguish between the quantum and
classical processes in the experiment.

The author thanks A. Bezryadin for useful comments.

\end{document}